\documentclass[aps,twocolumn,showpacs,preprintnumbers,prb]{revtex4}
\usepackage{graphicx}
\usepackage{dcolumn}
\usepackage{bm}
\usepackage{color}

\begin{document}

\title{Acoustic surface plasmons in the noble metals Cu, Ag, and Au}
\author{ V. M. Silkin,$^{1}$ J. M. Pitarke,$^{1,2}$ E. V. Chulkov,$^{1,3}$ and P. M. Echenique$^{1,3}$}
\affiliation{ $^1$Donostia International Physics Center (DIPC),
Manuel de Lardizabal Pasealekua,
E-20018 Donostia, Basque Country, Spain\\
$^2$Materia Kondentsatuaren Fisika Saila, Zientzi Fakultatea, Euskal
Herriko Unibertsitatea,
644 Posta kutxatila, E-48080 Bilbo, Basque Country, Spain\\
$^3$Materialen Fisika Saila, Kimika Fakultatea, Euskal Herriko
Unibertsitatea and Centro Mixto CSIC-UPV/EHU 1072 Posta kutxatila,
E-20080 Donostia, Basque Country, Spain}

\date{\today}

\begin{abstract}
We have performed self-consistent calculations of the dynamical
response of the (111) surface of the noble metals Cu, Ag, and Au.
Our results indicate that the partially occupied surface-state band
in these materials yields the existence of acoustic surface plasmons
with linear dispersion at small wave vectors. Here we demonstrate
that the sound velocity of these low-energy collective excitations,
which had already been predicted to exist in the case of Be(0001),
is dictated not only by the Fermi velocity of the two-dimensional
surface-state band but also by the nature of the decay and
penetration of the surface-state orbitals into the solid. Our
linewidth calculations indicate that acoustic surface plasmons
should be well defined in the energy range from zero to $\sim 400$
meV.
\end{abstract}


\maketitle

\section{Introduction}

During the last decades a variety of metal surfaces, such as
Be(0001) and the (111) surfaces of the noble metals Cu, Ag, and Au,
have become a testing ground for many experimental and theoretical
investigations.\cite{inrpp82,hu95,nissr02,ecbessr04} These surfaces
are known to support a partially occupied band of Shockley surface
states with energies near the Fermi level. Since these states are
strongly localized near the surface and disperse with momentum
parallel to the surface, they can be considered to form a quasi
two-dimensional (2D) surface-state band with a 2D Fermi energy
$\varepsilon_F^{2D}$ equal to the surface-state binding energy at
the $\bar\Gamma$ point.

In the absence of the three-dimensional (3D) substrate, partially
occupied Shockley surface states would support a 2D collective
oscillation, the energy of this plasmon being given by (unless
stated otherwise, atomic units are used, i.e.,
$e^2=\hbar=m_e=1$)\cite{stprl67,naheprl01}
\begin{equation}\label{one}
\omega_{2D}=\sqrt{2\pi\,n_{2D}q/m_{2D}},
\end{equation}
where $n_{2D}$ represents the density of occupied surface states:
$n_{2D}=\varepsilon_F^{2D}/\pi$, $q$ represents the magnitude of a
2D wave vector, and $m_{2D}$ is a 2D effective mass. Eq.~(\ref{one})
shows that at very long wavelengths plasmons in a 2D electron gas
have low energies; however, they do not affect electron-hole (e-h)
and phonon dynamics near the Fermi level, due to their square-root
dependence on the wave vector. Much more effective than ordinary 2D
plasmons in mediating, e.g., superconductivity would be the
so-called acoustic plasmons with sound-like long-wavelength
dispersion.\cite{matoap95}

Recently, it has been demonstrated that in the presence of the 3D
substrate the dynamical screening at the surface provides a
mechanism for the existence of a {\it new} acoustic collective mode,
whose energy exhibits a linear dependence on the 2D wave
vector.\cite{sigaepl04,pinaprb04} We refer to this mode as {\it
acoustic surface plasmon} (ASP), to distinguish it from the
conventional surface plasmon predicted by Ritchie.\cite{ripr57} The
energy of this latter plasmon is known to be
$\omega_s=\omega_p/\sqrt{2}$, where $\omega_p$ is the plasmon energy
of a homogeneous electron gas of density $n_0$: $\omega_p=(4\pi
n_0)^{1/2}$.

\begin{figure}
\includegraphics[width=0.55\linewidth,angle=270]{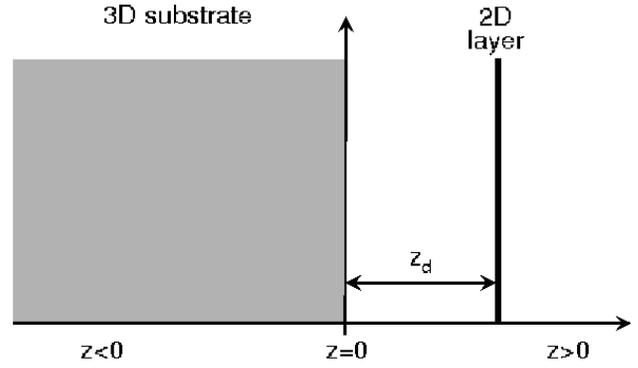}\\
\caption{Simplified model in which surface-state electrons comprise
a 2D sheet of interacting free electrons at $z=z_d$. All other
states of the semi-infinite metal are assumed to comprise a
plane-bounded 3D electron gas at $z\leq 0$. The metal surface is
located at $z=0$.} \label{fig1}
\end{figure}

In a simplified model in which surface-state electrons comprise a 2D
electron gas at $z=z_d$ (see Fig.~\ref{fig1}), while all other
states of the semi-infinite metal comprise a 3D substrate at $z\le
0$, one finds that both e-h and collective excitations occurring
within the 2D gas can be described with the use of an effective 2D
dielectric function, which in the random-phase approximation (RPA)
takes the form\cite{pinaprb04}
\begin{equation}\label{eff2}
\epsilon_{eff}(q,\omega)=1-W(z_d,z_d;q,\omega)\,\chi_{2D}^0(q,\omega),
\end{equation}
$W(z,z';q,\omega)$ being the 2D Fourier transform of the so-called
screened interaction in the presence of the 3D substrate
alone,\cite{ecpicp00} and $\chi_{2D}^0(q,\omega)$ being the
noninteracting density-response function of a 2D electron
gas.\cite{stprl67}

In the absence of the 3D substrate, $W(z,z';q,\omega)$ yields the 2D
Fourier transform of the bare Coulomb interaction and
$\epsilon_{eff}(q,\omega)$ coincides, therefore, with the RPA
dielectric function of a 2D electron gas, which in the
long-wavelength ($q\to 0$) limit has one single zero corresponding
to collective oscillations at $\omega=\omega_{2D}$.

In the presence of a 3D substrate, the long-wavelength limit of
$\epsilon_{eff}(q,\omega)$ has two zeros.\cite{note1} One zero
corresponds to a high-frequency oscillation of energy
$\omega^2=\omega_s^2+\omega_{2D}^2$ in which 2D and 3D electrons
oscillate in phase with one another. The other zero corresponds to a
low-frequency {\it acoustic} oscillation in which 2D and 3D
electrons oscillate out of phase. The energy of this low-frequency
mode is found to be of the form\cite{pinaprb04}
\begin{equation}\label{two}
\omega=\alpha\,v_F^{2D}\,q,
\end{equation}
where $v_F^{2D}$ represents the 2D Fermi velocity
\begin{equation}
v_F^{2D}=\sqrt{2\varepsilon_F^{2D}/m_{2D}},
\end{equation}
and
\begin{equation}\label{alpha}
\alpha=\sqrt{1+{\left[I(z_d)\right]^2\over\pi\left[\pi+2\,I(z_d)\right]}},
\end{equation}
with
\begin{equation}\label{limit2}
I(z_d)=\lim_{q\to 0}W(z_d,z_d;q,\alpha v_F^{2D}q).
\end{equation}
The coefficient $\alpha$ (whose value depends on the electron
density of the 3D substrate and increases with $z_d$) ranges from a
constant value (on the order of $1.3-1.6$ for metallic densities)
for a 2D sheet far inside the 3D susbtrate to the asymptotic value
$\sqrt{2z_d}$ (see Ref.~\onlinecite{note2}) for a 2D sheet far
outside the metal surface.

In this paper, we extend the self-consistent calculations of the
dynamical response of Be(001) reported in
Ref.~\onlinecite{sigaepl04} to the case of the (111) surface of the
noble metals Cu, Ag, and Au. Our results indicate that the partially
occupied surface-state band in these materials yields the existence
of acoustic surface plasmons whose energy is of the form of
Eq.~(\ref{two}), but with an $\alpha$ coefficient that is much
closer to unity than expected from the simplified model described
above. Furthermore, we demonstrate that the sound velocity
($v_s=\alpha v_F^{2D}$) of this low-energy collective excitation is
dictated not only by the Fermi velocity of the 2D surface-state band
but also by the nature of the decay and penetration of the
surface-state orbitals into the solid. We also investigate the width
of the corresponding plasmon peak, which dictates the lifetime of
this collective excitation.

\section{Theory}

In order to achieve a full description of the dynamical response of
real metal surfaces, we first consider the one-dimensional potential
of Ref.~\onlinecite{chsiss97}. This allows us to assume
translational invariance in the plane of the surface, which yields,
within linear-respone theory, the following expression for the
electron density induced by an external perturbation
$\phi^{ext}(z;q,\omega)$:
\begin{equation}\label{deltan}
\delta n(z;q,\omega)=\int
dz'\,\chi(z,z';q,\omega)\,\phi^{ext}(z';q,\omega),
\end{equation}
$\chi(z,z';q,\omega)$ representing the 2D Fourier transform of the
density-response function of our {\it interacting} many-electron
system.

The collective oscillations created by an external potential of the
form\cite{note3}
\begin{equation}\label{pot}
\phi^{ext}(z;q,\omega)=-(2\pi/q)\,{\rm e}^{qz}
\end{equation}
can be traced to the peaks of the imaginary part of the so-called
surface response function $g(q,\omega)$:\cite{fepss82,pezaprb85}
\begin{equation}\label{g}
g(q,\omega)=-{2\pi\over q}\int dz \int dz' \,{\rm e}^{q\,(z+z')}\,
\chi(z,z';q,\omega),
\end{equation}
which at $q=0$ exhibits a pole at the conventional surface plasmon
$\omega_s$.\cite{li97}

\subsection{Single-particle states}

The starting point of our calculations is a set of single-particle
states $\psi_{{\bf k},n}({\bf r})$ and energies $E_{{\bf k},n}$ of
the form
\begin{equation}
\psi_{{\bf k},n}({\bf r})={1\over\sqrt{A}}\,e^{i{\bf k}\cdot{\bf
r}_\parallel}\phi_n(z) \label{psimu}
\end{equation}
and
\begin{equation}
E_{{\bf k},n}=\frac{k^2}{2m_n}+\varepsilon_n, \label{emu}
\end{equation}
where ${\bf r} \equiv ({\bf r}_\parallel,z)$, $A$ is a normalization
area, and $\phi_n(z)$ and $\varepsilon_n$ are the eigenfunctions and
eigenvalues of a one-dimensional Schr\"odinger equation of the form
\begin{equation}
\left[-\frac{1}{2}\frac{d^2}{dz^2}+V_{MP}(z)\right]\phi_n(z)=\varepsilon_n
\phi_n(z), \label{Schrodinger}
\end{equation}
$V_{MP}$ being the model potential described in
Ref.~\onlinecite{chsiss97}. This potential reproduces the key
features of the surface band structure, which in the case of the
(111) surface of the noble metals are the presence of a band gap at
the center of the 2D Brillouin zone (2DBZ) and the existence of
Shockley and image states in it.

Alternatively, for a description of the screened interaction
$W(z,z';q,\omega)$ entering Eqs.~(\ref{eff2}) and (\ref{limit2})
(which accounts for the presence of 3D bulk states alone), the wave
functions $\phi_n(z)$ and $\varepsilon_n$ can be taken to be the
eigenfunctions and eigenvalues of a jellium Kohn-Sham Hamiltonian of
density-functional theory (DFT),\cite{hokopr64} which we evaluate in
the local-density approximation (LDA) with the parametrization of
Perdew and Zunger.\cite{pezuprb81}

In order to solve either Eq.~(\ref{Schrodinger}) or the jellium
Kohn-Sham equation of DFT, we consider a thick slab with a given
number of atomic layers and assume that the electron density
vanishes at a distance $z_0$ from either crystal edge.\cite{note4}
The one-dimensional wave functions $\phi_n(z)$ are then expanded in
a Fourier series of the form\cite{note5}
\begin{eqnarray}
&&\phi_n(z)={1\over \sqrt{d}}\,c^+_{n,0}+{\sqrt{2}\over
\sqrt{d}}\cr\cr &&\times\sum_{l=1}^{l_{max}}
\left[c^+_{n,l}\cos\left({2\pi l\over d}z\right)+
c^-_{n,l}\sin\left({2\pi l\over d}z\right)\right], \label{phi}
\end{eqnarray}
where the distance $d$ is given by the equation
\begin{equation}
d=N\,d_0+2z_0,
\end{equation}
$N$ and $d_0$ being the number of atomic layers and the interlayer
spacing, respectively; in the case of the (111) surfaces of the
noble metals, we take $N=81$, $z_0=10 d_0$, and $l_{max}$
corresponding to an energy of $150\,{\rm eV}$. Due to the symmetry
of the model potential entering Eq.~(\ref{Schrodinger}), the
eigenfunctions $\phi_n(z)$ are easily found to be either even
($c_{n,l}^-=0$) or odd ($c_{n,l}^+=0$).

\begin{figure}[tbp]
\centering
\includegraphics[scale=0.35,angle=270]{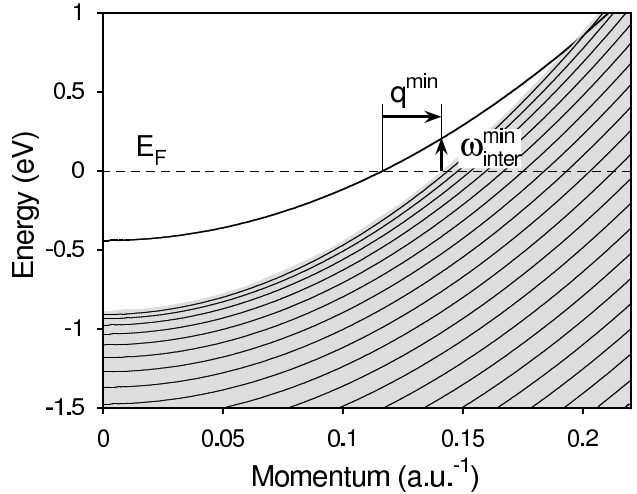}
\caption{Surface electronic structure of Cu(111). The thick solid
line is the energy $E_{{\bf k},n}=\varepsilon_F^{2D}+k^2/2m_{2D}$ of
the Shockley surface state versus the 2D momentum $k$, as obtained
with the measured value of the effective mass $m_{2D}$. The grey
area (with the upper limit at $E=\varepsilon_0+k^2/2m_0$) represents
the projected bulk band structure. The values of
$\varepsilon_F^{2D}$, $m_{2D}$, $\varepsilon_0$, and $m_0$
are reported in Table~\ref{table}. Solid lines correspond to the
bulk states that we have obtained by employing a slab with 81 atomic
layers to simulate the semi-infinite solid. We have used the
following parameters in the description of the model potential
$V_{MP}$ entering Eq.~(\ref{Schrodinger}): $A_{10}=-11.805$,
$A_1=5.14$, $A_2=4.4204$, and $\beta=2.8508$ (for a description of
the model potential see Ref.~\onlinecite{chsiss97}). Due to the
presence of the band gap, for optical ($q=0$) transitions to occur
from an occupied 3D bulk state to an unoccupied 2D surface state the
minimum energy $\omega_{\rm inter}^{\rm min}$ is required, which
decreases as the momentum transfer $q$ increases. For $q$ larger
than $q^{\rm min}$, transitions from occupied (unoccupied) 3D bulk
states to unoccupied(occupied) surface states can occur at arbitrary
values of the energy transfer $\omega$. } \label{fig2}
\end{figure}

Figure~\ref{fig2} shows the energies $E_{{\bf k},n}$ that we have
obtained from Eq.~(\ref{emu}) \textcolor[rgb]{0.50,0.00,0.00} {for}
Cu(111) by solving Eq.~(\ref{Schrodinger}) as described above and
using the experimental values of the effective masses $m_n$
\textcolor[rgb]{0.50,0.00,0.00} {of all bulk and surface surface
states}. Electronic structures for Ag(111) and Au(111) are similar.
The corresponding parameters are reported in Table~\ref{table}.

\subsection{Noninteracting density-response function}

Once we have an accurate description of the single-particle orbitals
$\phi_n(z)$ and energies $\varepsilon_n$, we evaluate the 2D Fourier
transform $\chi^0(z,z';q,\omega)$ of noninteracting electrons moving
in either the model potential $V_{MP}(z)$ or the jellium effective
Kohn-Sham potential of DFT:
\begin{eqnarray}
&&\chi^{0}(z,z';q,\omega)=\frac{2}{A} \mathrel{\mathop{\sum
}\limits_{n,n'}}\phi_{n}(z)\phi_{n'}(z)\phi_{n}(z')\phi
_{n'}(z')\cr\cr &&\times \mathop{\sum}\limits_{{\bf k}}
\frac{f_{{\bf k},n}-f_{{\bf k}+{\bf q},n'}}{E_{{\bf k},n}- E_{{\bf
k}+{\bf q},n'}+\omega+i\eta}. \label{chi0zz1}
\end{eqnarray}
Here, the sum over $n$ and $n'$ includes both occupied and
unoccupied states, $\eta$ is a positive infinitesimal, and $f_{{\bf
k},n}$ are Fermi factors, which at zero temperature are simply given
by the Heaviside step function
\begin{equation}
f_{{\bf k},n}=\Theta(\varepsilon_F-E_{{\bf k},n}),
\end{equation}
$\varepsilon_F$ being the Fermi energy of the solid. In particular,
the noninteracting density-response function of the quasi-2D band of
occupied Shockley states in the absence of the 3D substrate can be
obtained from Eq.~(\ref{chi0zz1}) by omitting all bulk states in the
sum over $n$ and $n'$.

Introducing the one-dimensional wave functions of Eq.~(\ref{phi})
into Eq.~(\ref{chi0zz1}), one finds the following Fourier
representation of the noninteracting density-response function (see
Appendix A):\cite{note7}
\begin{widetext}
\begin{equation}
\chi^0(z,z';q,\omega)=\sum_{n=0}^\infty\,\sum_{n'=0}^\infty
\chi^{0,+}_{n,n'}(q,\omega)\cos\left({2\pi n\over d}z\right)
\cos\left({2\pi n'\over d}z'\right)+
\sum_{n=1}^\infty\,\sum_{n'=1}^\infty
\chi^{0,-}_{n,n'}(q,\omega)\sin\left({2\pi n\over d}z\right)
\sin\left({2\pi n'\over d}z'\right). \label{chi0gg}
\end{equation}
\end{widetext}

\subsection{Interacting density-response function}

In the framework of the RPA,\cite{fewa64} the 2D Fourier transform
$\chi(z,z';q,\omega)$ of the density-response function of an
interacting many-electron system is obtained by solving the
following integral equation
\begin{eqnarray}
\chi(z,z';q,\omega)&=&\chi^0(z,z';q,\omega) +\int {\rm d}z_1 \int
{\rm d}z_2
\chi^0(z,z_1;q,\omega)  \nonumber \\
&\times& v(z_1,z_2;q)\,\chi(z_2,z';q,\omega), \label{chichi0}
\end{eqnarray}
the ingredients of this equation being the 2D Fourier transforms
$\chi^0(z,z';q,\omega)$ and $v(z,z';q)$ of the noninteracting
density-response function and the bare Coulomb interaction,
respectively. All quantities entering Eq.~(\ref{chichi0}) can be
represented in the form of Eq.~(\ref{chi0gg}), which yields the
following matrix equation for the coefficients
$\chi^{\pm}_{m,n}(q,\omega)$:
\begin{eqnarray}
\chi^{\pm}_{n,n'}(q,\omega)&=&\chi^{0,\pm}_{n,n'}(q,\omega)+
\sum_{n'',n'''}\chi^{0,\pm}_{n,n''}(q,\omega)\cr\cr
&\times&v_{n'',n'''}(q)\,\chi^{\pm}_{n''',n'}(q,\omega),
\end{eqnarray}
$v_{n'',n'''}(q)$ being the corresponding coefficients of the bare
Coulomb interaction $v(z,z';q)$.

\begin{table*}
\caption{\label{table} 2D Fermi energy ($\varepsilon_F^{2D}$),
effective mass ($m_{2D}$), and Fermi velocity ($v_F^{2D}$) of the
Shockley surface-state band in the (111) surface of the noble metals
Cu, Ag, and Au. $\varepsilon_0$ represents the energy of the bottom
of the gap at the $\bar\Gamma$ point. $m_0$ represents the effective
mass of the upper bulk states at the bottom of the gap. The minimum
energy transfer $\omega_{\rm inter}^{\rm min}$ and momentum transfer
$q^{\rm min}$ are those defined in Fig.~2. Also represented in this
table are the values of the parameter $\alpha$ that we have obtained
from our full self-consistent calculations of the surface-response
function of Eq.~(\ref{g}) and from Eq.~(\ref{alpha}) with $z_d<<0$.}
\begin{ruledtabular}
\begin{tabular}{crcclccccc}
& $\varepsilon_F^{2D}$ (meV) & $m_{2D}$ & $v_F^{2D}$ &
$\varepsilon_0$ (eV) & $m_0$ & $\omega_{\rm inter}^{\rm min}$ (meV)
& $q^{\rm min}$ & $\alpha$ & $\alpha$
           [Eq.~(\ref{alpha})] \\
\hline
Cu & $440$ & $0.42$ & $0.277$ & $-0.89$ & $0.31$ & $217$ & $0.026$ & $1.053$ & $1.38$ \\
Ag & $67$  & $0.44$ & $0.106$ & $-0.4 $ & $0.25$ & $160$ & $0.039$ & $1.042$ & $1.41$ \\
Au & $475$ & $0.28$ & $0.353$ & $-1.0 $ & $0.21$ & $275$ & $0.025$ & $1.032$ & $1.41$ \\
\end{tabular}
\end{ruledtabular}
\end{table*}

\section{Results and discussion}

\begin{figure}[tbp]
\centering
\includegraphics[scale=0.48,angle=0]{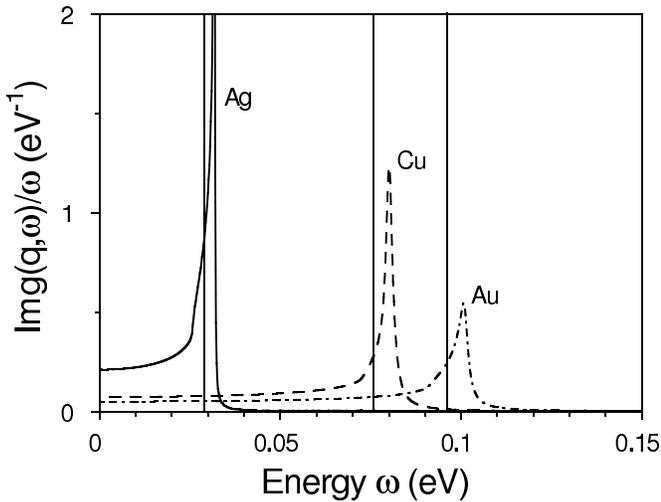}
\caption{Energy-loss function ${\rm Im} g(q,\omega)/\omega$ of the
(111) surfaces of the noble metals Cu, Ag, and Au, shown by solid,
dashed, and dashed dotted lines, respectively, versus the excitation
energy $\omega$, as obtained from Eq.~(\ref{g}) for $q=0.01$ and
$\eta=1$ meV. The vertical solid lines are located at the energies
$\omega=v_F^{2D}\,q$, which would correspond to Eq.~(\ref{two}) with
$\alpha=1$.} \label{fig3}
\end{figure}

\subsection{ASP dispersion}

Figure~\ref{fig3} shows the energy-loss function ${\rm
Im}g(q,\omega)$ of the (111) surfaces of the noble metals Cu, Ag,
and Au, as obtained from Eq.~(\ref{g}) for $q=0.01$. This figure
shows the presence of a low-energy collective excitation, whose
energy is of the form of Eq.~(\ref{two}) but with an $\alpha$
coefficient that is close to unity. Furthermore, we have carried out
calculations of ${\rm Im}g(q,\omega)$ for several low values of $q$
and we have found that this low-energy collective excitation is
indeed an acoustic surface plasmon with linear dispersion, as shown
in Fig. 4 for the case of Cu(111).

\begin{figure}[tbp]
\centering
\includegraphics[scale=0.35,angle=270]{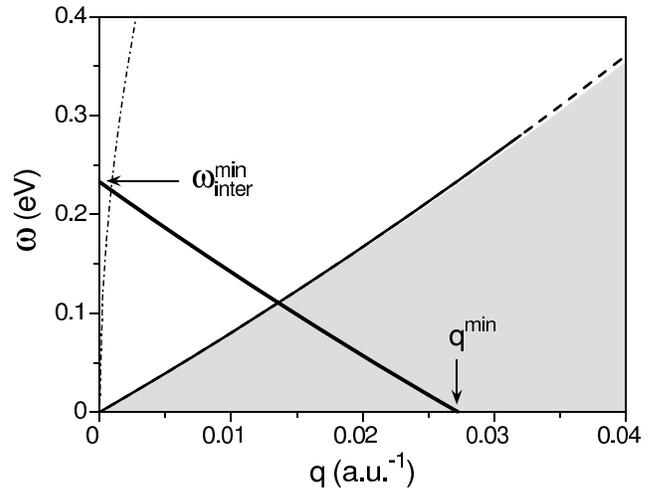}
\caption{The solid line shows the energy of a well-defined acoustic
surface plasmon of Cu(111), as obtained from the maxima of our
calculated surface-loss function ${\rm Im}g(q,\omega)$. The dashed
line represents the energy of an acoustic surface plasmon whose
linewidth starts to be considerable due to the presence of interband
transitions. The dashed dotted line is the plasmon dispersion of a
2D electron gas in the absence of the 3D system. The grey area
indicates the region of the $(q,\omega)$ plane (with the upper limit
at $\omega_{2D}^{\rm up}=v_F^{2D}q+q^2/2m_{2D}$) where e-h pairs can
be created within the 2D Shockley band of Cu(111).
The area below the thick solid line corresponds to the region of
momentum space where transitions between 3D and 2D states cannot
occur. The quantities $\omega_{\rm inter}^{\rm min}$ and $q^{\rm
min}$ are determined from the surface band structure of
Fig.~\ref{fig2}.} \label{fig4}
\end{figure}

In Fig.~\ref{fig4}, we show the energy of the acoustic surface
plasmon of Cu(111) versus $q$ (solid line), as derived from the
maxima of our calculated ${\rm Im}g(q,\omega)$, together with the
well-defined plasmon energies that we obtain when only the
surface-state band is considered in the evaluation of the
noninteracting density-response function of Eq.~(\ref{chi0zz1})
(dashed dotted line). While the plasmon energies of electrons in the
isolated surface-state band nicely reproduce in the long-wavelength
($q\to 0$) limit the conventional 2D plasmon dispersion
$\omega_{2D}$ of Eq.~(\ref{one}), the combination of this
surface-state band with the underlying 3D system yields a {\it new}
distinct mode whose energy lies just above the upper edge
$\omega_{2D}^{\rm up}=v_F^{2D}q+q^2/2m_{2D}$ of the 2D e-h pair
continuum, as occurs in the case of Be.\cite{sigaepl04} Furthermore,
Fig.~\ref{fig4} shows that in the long-wavelength ($q\to 0$) limit
the energy of the acoustic surface plasmon in Cu(111) is of the form
of Eq.~(\ref{two}) but with an $\alpha$ coefficient that is
considerably closer to unity than expected from Eq.~(\ref{alpha})
(see Table~\ref{table}). This discrepancy can be originated in (i)
the absence in the simplified model leading to Eqs.~(\ref{two}) and
(\ref{alpha}) of transitions between 2D and 3D states, and (ii) the
nature of the decay and penetration of the surface-state orbitals
into the solid.

\begin{figure}[tbp]
\centering
\includegraphics[scale=0.48,angle=270]{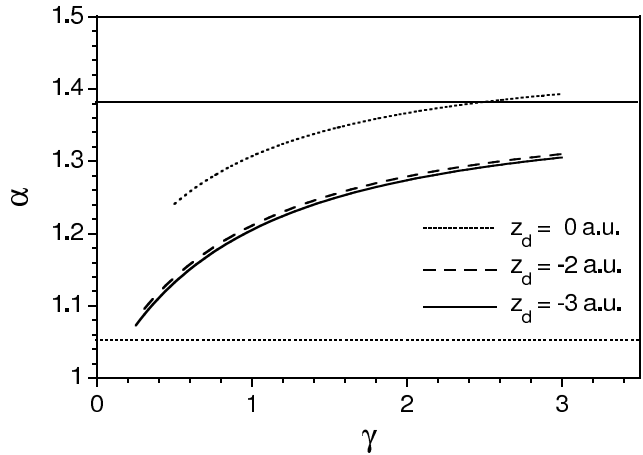}
\caption{In a simplified model in which the wave function of
surface-state electrons decays exponentially away from the $z=z_d$
plane (see Fig.~\ref{fig1}) with a decay constant $\gamma$, acoustic
surface plasmons are found to exist whose energy is of the form of
Eq.~(\ref{two}) with the coefficient $\alpha$ of Eq.~(\ref{alpha})
replaced by the $\gamma$-dependent $\alpha$ coefficient that we have
presented in this figure in the case of Cu(111) for various values
of $z_d$. The horizontal solid line corresponds to the coefficient
$\alpha$ of Eq.~(\ref{alpha}) with $z_d<<0$. The horizontal dotted
line represents the coefficient $\alpha$ derived from our full
self-consistent calculation that treats bulk and surface states on
the same footing. As $z_d$ is shifted from the interior of the solid
towards the vacuum, the coefficient $\alpha$ increases, in agreement
with Ref.~\onlinecite{pinaprb04}.} \label{fig5}
\end{figure}

\begin{figure}[tbp]
\centering
\includegraphics[scale=0.36,angle=270]{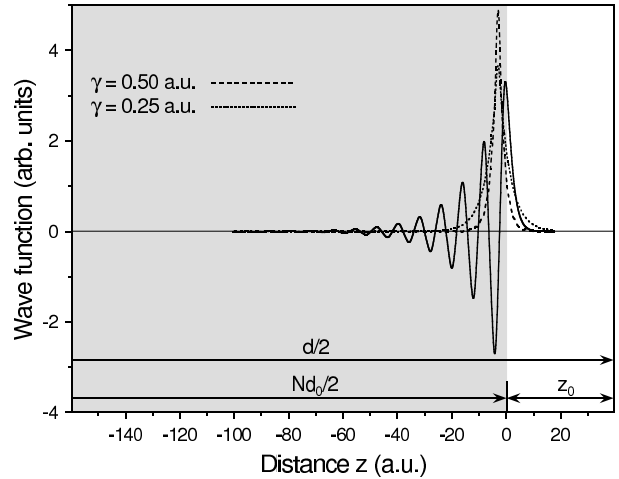}
\caption{The solid line is the wave function $\phi(z)$ of the
occupied Shockley surface state of Cu(111) at the $\bar\Gamma$
point, as obtained by solving Eq.~(\ref{Schrodinger}). The dashed
and dotted lines show wave functions of the form
$\phi(z)\sim\exp(-\gamma|z-z_d|)$ with $z_d=-3$ a.u. and two
different values of the decay constant: $\gamma=0.5\,a_0^{-1}$
(dashed line) and $\gamma=0.25\,a_0^{-1}$ (dotted line). $a_0$ is
the Bohr radius: $a_0=0.529\,{\rm\AA}$.} \label{fig6}
\end{figure}

In order to investigate the origin of the small differences between
the plasmon energies obtained here and those expected from
Eqs.~(\ref{two}) and (\ref{alpha}), we have carried out calculations
of ${\rm Im}g(q,\omega)$ along the lines of the simplified model
leading to Eqs.~(\ref{two}) and (\ref{alpha}) (see Fig.~\ref{fig1})
but with the 2D electron gas of Fig.~\ref{fig1} replaced by a more
realistic quasi-2D gas of electrons described by a wave function
that decays exponentially away from the $z=z_d$ plane with a decay
constant $\gamma$. We have found that an acoustic surface plasmon is
present whose energy is indeed of the form of Eq.~(\ref{two}) but
with an $\alpha$ coefficient that strongly depends on the decay
constant $\gamma$, as shown in Fig.~\ref{fig5}. This figure
demonstrates that while in the limit as $\gamma\to\infty$ (where the
quasi-2D electron gas is indeed an ideal 2D sheet) the coefficient
$\alpha$ approaches the value expected from Eq.~(\ref{alpha})
(horizontal solid line), as $\gamma$ decreases the dispersion of the
acoustic surface plasmon approaches (for all negative values of
$z_d$) the more realistic situation where $\alpha$ is close to unity
(horizontal dotted line). A comparison between the model wave
functions that we have used in this calculation and the actual
surface-state wave functions that are involved in the full
calculation of Figs.~\ref{fig3} and \ref{fig4} is presented in
Fig.~\ref{fig6}. Although the {\it model} wave functions do not
reproduce the actual shape of the surface-state wave function, a
finite penetration of the model surface-state wave functions into
the solid allows the formation of an acoustic surface plasmon whose
sound velocity is very close to the Fermi velocity of the 2D
surface-state band ($\alpha\sim 1$), as predicted by our more
realistic calculation. Indeed, the finite penetration of the
surface-state wave function into the solid provides a more complete
screening of the quasi-2D collective excitations by the surrounding
3D substrate, which brings the acoustic surface plasmon closer to
the upper edge of the 2D e-h pair continuum ($\alpha\to
1$).\cite{notenew}

\begin{figure}[tbp]
\centering
\includegraphics[scale=0.35,angle=270]{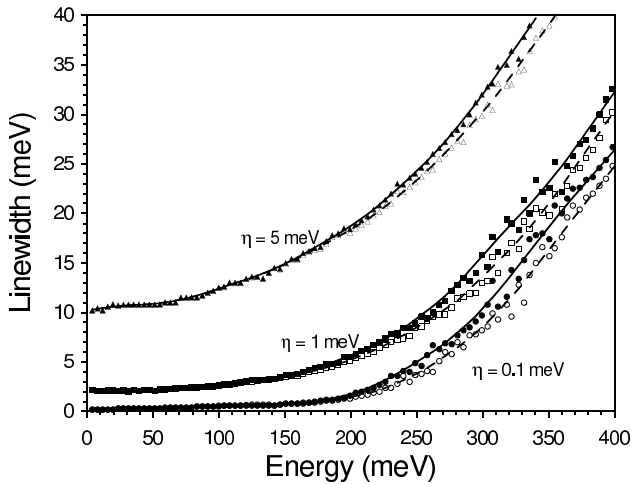}
\caption{Solid and open symbols represent the width at half maximum
of acoustic surface plasmons in Cu(111) versus the plasmon energy,
as obtained from the imaginary part of the surface-response function
$g(q,\omega)$ of Eq.~(\ref{g}) with (solid symbols) and without
(open symbols) inclusion of transitions between 2D and 3D states in
the evaluations of the noninteracting density-response function of
Eq.~(\ref{chi0zz1}) and for various values of the parameter $\eta$:
0.1, 1, and 5 meV. The solid and dashed lines represent fits from
the solid and open symbols, respectively. For this surface, the
threshold for interband transitions between 2D and 3D states occurs
at $\sim 110$ meV (see Fig.~\ref{fig4}).} \label{fig7}
\end{figure}

\subsection{ASP linewidth}

Finally, we have carried out lifetime calculations of the acoustic
surface plasmon, as derived from the width of the imaginary part of
the full surface-response function of Eq.~(\ref{g}). Fig.~\ref{fig7}
shows the results we have obtained for Cu(111) with (solid symbols
and lines) and without (open symbols and dashed lines) inclusion of
transitions between 2D and 3D states in the evaluation of the
noninteracting density-response function of Eq.~(\ref{chi0zz1}) and
for three different values of the parameter $\eta$. Plasmon decay
can occur by exciting e-h pairs either through transitions between
2D and 3D states, which would not be present in the model leading to
Eqs.~(\ref{two}) and (\ref{alpha}), or through transitions within
the 3D continuum of bulk states.\cite{note8} At small energies below
the threshold of $\omega\sim 110\,{\rm meV}$ (see Fig.~\ref{fig4}),
where acoustic surface plasmons can only decay by exciting e-h pairs
within the 3D continuum of bulk states (the solid and dashed lines
of Fig.~\ref{fig7} coincide), the linewidth is entirely determined
by the choice of the parameter $\eta$, showing that at these low
energies the impact of intraband transitions between 3D bulk states
is negligibly small. As the plasmon energy increases, there is a
small contribution to the plasmon linewidth from transitions between
2D and 3D states (the difference between solid and dashed
lines)\cite{notenew2} and an increasing contribution from intraband
3D transitions yielding a finite linewidth which for $\eta<0.1\,{\rm
eV}$ is not sensitive to the precise value of $\eta$ employed, but
which still allows the formation of a well-defined acoustic-surface
collective excitation for plasmon energies at least up to
$\sim400\,{\rm meV}$. A similar behavior is observed in the case of
Ag and Au.

\section{Summary and conclusions}

We have carried out self-consistent calculations of the surface-loss
function of the (111) surfaces of the noble metals Cu, Ag, and Au,
by considering a one-dimensional potential that describes the main
features of the surface band structure. We have found that the
partially occupied surface-state band in these materials yields the
existence of acoustic surface plasmons, as it had already been
demonstrated to occur in the case of Be(0001).\cite{sigaepl04} The
energy of these collective excitations has been shown to exhibit
linear dispersion at small wave vectors, with the sound velocity
being very close to the Fermi velocity of the 2D surface-state band
and considerably smaller than expected from a simplified model in
which surface-state electrons comprise a 2D electron gas while all
other states of the semi-infinite metal comprise a 3D substrate.

The origin of the differences between the plasmon energies obtained
here and those expected from simplified models has been investigated
by performing simplified calculations with the ideal 2D sheet of
Shockley electrons replaced by a more realistic quasi-2D gas of
electrons whose wave functions decay exponentially away from the
surface. These calculations have been found to yield an acoustic
surface plasmon whose energy is linear in the magnitude of the wave
vector, the sound velocity being dictated not only by the Fermi
velocity of the 2D surface-state band but also by the nature of the
decay and penetration of the surface-state orbitals into the solid.
With an appropriate choice of the exponential decay of surface-state
wave functions these simplified calculations (which do not account
for transitions between 2D and 3D states) accurately account for the
energy dispersion of acoustic surface plasmons, which indicates that
the impact of interband transitions between 2D and 3D states on the
ASP's energy dispersion is negligible.

We have also carried out self-consistent calculations of the
linewidth of acoustic surface plasmons in the (111) surfaces of the
noble metals. We have found that while the impact of interband
transitions between 2D and 3D states is small, intraband transitions
between 3D bulk states contribute considerably to the finite
linewidth of acoustic surface plasmons, which are found to represent
a well-defined acoustic collective excitation for plasmon energies
at least up to $\sim 400\,{\rm meV}$.

Finally, we note that as in the case of conventional surface
plasmons, acoustic surface plasmons should also be expected to be
excited by light, as discussed recently.\cite{optics}

\section{Acknowledgments}

We gratefully acknowledge partial support by the University of the
Basque Country, the Basque Unibertsitate eta Ikerketa Saila, the
Spanish Ministerio de Ciencia y Tecnolog\'\i a, and the EC 6th
framework Network of Excellence NANOQUANTA (NMP4-CT-2004-500198).

\appendix
\section{}

Here we give explicit expressions for the coefficients
$\chi_{n,n'}^{0,\pm}(q,\omega)$ entering the expansion of the
noninteracting density-response function of Eq.~(\ref{chi0gg}), as
obtained by introducing the one-dimensional wave functions of
Eq.~(\ref{phi}) into Eq.~(\ref{chi0zz1}). Replacing the sum over
{\bf k} in Eq.~(\ref{chi0zz1}) by an integral, we find
\begin{widetext}
\begin{equation}
\chi^{0,+}_{n,n'}(q,\omega)=\frac{\delta_n
\delta_{n'}}{d^2}\left\{\mathrel{\mathop{\sum}
\limits_{l_{even}}^{occ}}
\mathrel{\mathop{\sum}\limits_{l'_{even}}^{\infty}}\,
F_{l,l'}(q,\omega) G^{++}_{n;l,l'} G^{++}_{n';l,l'}+
\mathrel{\mathop{\sum}\limits_{l^{ }_{odd}}^{occ}}
\mathrel{\mathop{\sum}\limits_{l'_{odd}}^{\infty}}\,
F_{l,l'}(q,\omega) G^{--}_{n;l,l'} G^{--}_{n';l,l'} \right\}
\label{chi0C}
\end{equation}
and
\begin{equation}
\chi^{0,-}_{n,n'}(q,\omega)=\frac{4}{d^2}\left\{
\mathrel{\mathop{\sum}\limits_{l^{ }_{even}}^{occ}}
\mathrel{\mathop{\sum}\limits_{l'_{odd}}^{\infty}}\,
F_{l,l'}(q,\omega) G^{+-}_{n;l,l'} G^{+-}_{n';l,l'} +
\mathrel{\mathop{\sum}\limits_{l^{ }_{odd}}^{occ}}
\mathrel{\mathop{\sum}\limits_{l'_{even}}^{\infty}}\,
F_{l,l'}(q,\omega) G^{-+}_{n;l,l'} G^{-+}_{n';l,l'}\right\},
\label{chi0S}
\end{equation}
where
\begin{equation}
\delta_n=\cases{1,&for $n=0$\cr\cr 2,& for $n\ge1$}
\end{equation}
and
\begin{eqnarray}
 F_{l,l'}(q,\omega)=\int{\rm d}k\,k
\left[F^+_{l,l'}(q,k,\omega)-F^-_{l,l'}(q,k,\omega)\right],
\label{chi0zz}
\end{eqnarray}
with
\begin{eqnarray}
F^{\pm}_{l,l'}(q,k,\omega)=
\theta\left(\varepsilon_F-\epsilon_l-\frac{k^2}{2m_l}\right)
\left[\frac{q^2k^2}{m^2_{l'}}
-\left(\frac{q^2}{2m_{l'}}+\epsilon_{l'}-\epsilon_l+\frac{k^2}{2m_{l'}}-
\frac{k^2}{2m_{l}} \pm \omega \pm i\eta\right)^2\right]^{-1/2},
\label{fpm}
\end{eqnarray}
\begin{eqnarray}
G^{++}_{n;l,l'} = c^{+}_{l,0}c^{+}_{l',0}\delta_{n,0} &&+
\frac{1}{\sqrt{2}}\mathrel{\mathop{\sum}\limits_{n'\neq0}}
(c^{+}_{l,n'}c^{+}_{l',0}
+c^{+}_{l,0}c^{+}_{l',n'})\delta_{n,n'} \nonumber \\
&&+ \frac{1}{2}\mathrel{\mathop{\sum}\limits_{n'\neq0}}
\mathrel{\mathop{\sum}\limits_{n''\neq0}}c^{+}_{l,n'}c^{+}_{l',n''}
(\delta_{n+n'',n'}+\delta_{n+n',n''}+\delta_{n'+n'',n}),
 \label{gpp}
\end{eqnarray}
\begin{equation}
G^{--}_{n;l,l'} =
\frac{1}{2}\mathrel{\mathop{\sum}\limits_{n'\neq0}}
\mathrel{\mathop{\sum}\limits_{n''\neq0}}c^{-}_{l,n'}c^{-}_{l',n''}
(\delta_{n+n'',n'}+\delta_{n+n',n''}-\delta_{n'+n'',n}),
\label{gpp2}
\end{equation}
\begin{equation}
G^{+-}_{n;l,l'} = \frac{1}{\sqrt{2}}c^{+}_{l,0}c^{-}_{l',n} +
\frac{1}{2}\mathrel{\mathop{\sum}\limits_{n'\neq0}}
\mathrel{\mathop{\sum}\limits_{n''\neq0}}c^{+}_{l,n'}c^{-}_{l',n''}
(-\delta_{n+n'',n'}+\delta_{n+n',n''}+\delta_{n'+n'',n}),
 \label{gpp3}
\end{equation}
and
\begin{equation}
G^{-+}_{n;l,l'} = \frac{1}{\sqrt{2}}c^{-}_{l,n}c^{+}_{l',0} +
\frac{1}{2}\mathrel{\mathop{\sum}\limits_{n'\neq0}}
\mathrel{\mathop{\sum}\limits_{n''\neq0}}c^{-}_{l,n'}c^{+}_{l',n''}
(\delta_{n+n'',n'}-\delta_{n+n',n''}+\delta_{n'+n'',n}).
 \label{gpp4}
\end{equation}
\end{widetext}


\begin{thebibliography}{100}
\bibitem{note3} This would be the potential created by, e.g., electrons moving far
from the metal in the vacuum side of the surface.
\bibitem{note4} The crystal edge is taken to be half an interlayer spacing beyond the last atomic layer.
\bibitem{note5} The vacuum distance $z_0$ and the upper limit $l_{max}$ are chosen
sufficiently large for the physical results to be insensitive to the
precise value employed.
\bibitem{notenew} If the 2D collective excitations were completely screened by the 3D substrate,
no low-energy acoustic mode would exist, as shown in Ref.~\onlinecite{pinaprb04}.
\bibitem{note8} Since the 3D Fermi velocity $v_F^{3D}$ of the noble metals is
larger than the Fermi velocity $v_F^{2D}$ of the corresponding 2D
surface-state band, acoustic surface plasmons can always decay by
exciting e-h pairs within the 3D continuum of bulk states.
\bibitem{notenew2} This contribution to the plasmon linewidth is small due to the small
coupling between 2D and 3D orbitals.
\end{thebibliography}
\end{document}